\documentclass[preprint,english,showpacs,12pt,floatfix]{revtex4-1}
\usepackage{babel,amsmath,amssymb,dcolumn}
\usepackage{hyperref}
\usepackage{amsfonts}
\usepackage{amssymb}
\usepackage{graphicx}
\usepackage{graphics}
\usepackage{latexsym}
\usepackage{dcolumn}
\usepackage{epsfig}
\usepackage{subfigure}
\usepackage{array}
\usepackage{multirow}
\usepackage[retainorgcmds]{IEEEtrantools}
\allowdisplaybreaks[1]
\begin{document}

\newcommand{\vp}{\varphi}
\newcommand{\nn}{\nonumber\\}
\newcommand{\beq}{\begin{equation}}
\newcommand{\eeq}{\end{equation}}
\newcommand{\bed}{\begin{displaymath}}
\newcommand{\eed}{\end{displaymath}}
\def\bea{\begin{eqnarray}}
\def\eea{\end{eqnarray}}
\newcommand{\veps}{\varepsilon}
\newcommand{\nablasl}{{\slash \negthinspace \negthinspace \negthinspace \negthinspace  \nabla}}
\newcommand{\om}{\omega}

\newcommand{\Dsl}{{\slash \negthinspace \negthinspace \negthinspace \negthinspace  D}}
\newcommand{\tDsl}{{\tilde \Dsl}}
\newcommand{\tnablasl}{{\tilde \nablasl}}
\title{Vacuum polarization of the quantized massive scalar field in the global monopole spacetime II: the renormalized quantum stress energy tensor}

\author{Owen Pavel Fern\'{a}ndez Piedra$^{1, 2}$ \\
\textit{$^1$ Departamento de F\'isica, Divisi\'on de Ciencias e Ingenier\'ias, Universidad de Guanajuato, Campus Le\'on, Loma del Bosque N0. 103, Col. Lomas del Campestre, CP 37150, Le\'on, Guanajuato, M\'exico.}\\
\textit{$^2$Grupo de Estudios Avanzados, Universidad de Cienfuegos, Carretera a Rodas, Cuatro Caminos, s/n. Cienfuegos, Cuba.}}
\email{opavelfp2006@gmail.com}

\begin{abstract}
This paper is devoted to the construction of the renormalized quantum stress energy tensor $\left<T_{\mu}^{\nu}\right>_{ren}$ for a massive scalar field with arbitrary coupling to the gravitational field of a pointlike global monopole, using the Schwinger-DeWitt approximation, up to second order in the inverse mass $\mu$ of the field. The given stress energy tensor is constructed by functional differentiation with respect to the metric tensor of the one-loop effective action of sufficiently massive scalar field, such that the Compton length of the quantum field is much less than the characteristic radius of the curvature of the background geometry. The results are obtained for a general curvature coupling parameter $\xi$, and specified to the more physical cases of minimal and conformal coupling, showing that in this specific cases, the quantum massive scalar field in the global monopole spacetime violates all the pointwise energy conditions.
\end{abstract}
\date{\today}
\maketitle

\section{Introduction}
Quantum Field theory in curved spacetime is a well-established branch of modern physics, which has allowed the achievement of novel results since Hawking's discovery of black hole radiation \cite{Hawking}. Within this framework, we consider the quantum dynamics of fields in a gravitational background, considered as a classical external field. That is, all matter fields are considered using quantum field theory, with the only exception of the external gravitational field, that remains satisfying the classical Einstein field equations of General Relativity \cite{DeWitt,birrel}.

In this context, an important role is played by the quantum stress energy tensor \(\langle T_{\mu}^{\nu}\rangle\) of the quantum field, which is used as a source in the so called semiclassical Einstein's equations to take a look at the quantum corrections to the background geometry caused by the quantization of matter fields \cite{birrel,york,lousto-sanchez}. For this reason, it is very useful to have explicit analytical expressions for the renormalized stress tensor. This quantum stress tensor and the expectation value of the field fluctuation \(\langle\varphi^{2}\rangle_{ren}\) of a quantum field \(\varphi\) are the main objects to be determined from quantum field theory in curved spacetime.

The exact determination of \(\langle T_{\mu}^{\nu}\rangle\) for a generic spacetime is very cumbersome, and some techniques have been developed and applied to this problem, including numerical ones \cite{candelas,koffman,frolov-zelnikov,avramidi,AHS,matyjasek1,matyjasek,berej-matyjasek,owen1,owen2,Folacci,FT}. However, for the case of massive fields, there exist a method, called the Schwinger-DeWitt effective action approach, in which one assumes that the vacuum polarization effects can be separated from the particle creation, for masses of the fields sufficiently large. This method allow us to obtain approximate analytical expressions for the one-loop quantum effective action as an expansion in the square of the inverse mass of the quantum field. From the effective action, the quantum stress energy tensor can be calculated by functional differentiation with respect to the metric. This approach, based on proper time expansion of the Green's function of the dynamical operator that describes the evolution of the quantum field, can be used to investigate effects like the vacuum polarization of massive fields in curved backgrounds, whenever the Compton's wavelenght of the field is less than the characteristic radius of curvature \cite{DeWitt,frolov-zelnikov,avramidi,AHS,matyjasek1,matyjasek,berej-matyjasek,popov,owen1,owen2,Folacci}.

In a previous paper \cite{owenmonopole1} we took the first step in the investigation of vacuum polarization effects of quantum massive scalar field with arbitrary coupling to the gravitational field of a point-like global monopole, using the Schwinger-DeWitt technique to obtain analytical expressions for the field fluctuation \(\langle\varphi^{2}\rangle_{ren}\) in this background spacetime. We used the simple model, discovered by Barriola and Vilenkin \cite{barriola-vilenkin}, which leads to global monopoles as heavy objects appeared in the early universe as a result of a phase transition of a self-coupled scalar field triplet whose original global $O(3)$ symmetry is spontaneously broken to $U(1)$. In this systems, the scalar field plays the role of order parameter which is nonzero outside the monopole's core, where it is concentrated the main part of the monopole's energy.

Previous works that consider quantum fields in global monopole systems includes the analysis of massless scalar fields \cite{hiscock,mazzitely-lousto,bezerra1}, and the calculation of the quantum stress energy tensor for a massless spinor field \cite{bezerra2,bezerra3}.

In \cite{owenmonopole1}, we construct various approximations for $<\phi^{2}>$, each one proportional to the coincident limit of the Hadamard-DeWitt coeficcient $\left[a_{k}\right]$, starting from the leading term, proportional to $\left[a_{2}\right]$, up to the next to next to next to leading term, that include the coincident limits of coefficients up to $\left[a_{5}\right]$. In terms of the mass $\mu$ of the quantum scalar field, the leading approximation lead to $<\phi^{2}>$ proportional to $\mu^{-2}$, whereas the higher order approximation involve powers $\mu^{-8}$.  We also find the trace of the renormalized stress energy tensor for the quantized field in the leading approximation, using the existing relationship between this magnitude, the trace anomaly and the field fluctuation.

The results obtained in \cite{owenmonopole1} for the field fluctuations of the quantized massive scalar field in the global monopole background shows that taking into account higher order terms substantially improve the approximation, for which we concluded that for this spacetimes, we need to use the next to next to next to leading term to obtain a good description of the vacuum polarization.

This situation was in contrast with that obtained for other spacetimes with spherical symmetry, as the one describing a Reissner-Nordstrom black hole, for which previous studies showed that the next to leading term, proportional to $\left[a_{3}\right]$, provides a reasonable good approximation \cite{matyjasek4}.

In this paper we continue the study of vacuum polarization effects in the spacetime of the global monopole. Using the Schwinger-DeWitt approach, we construct an analytic expression of the four dimensional renormalized quantum effective action for a quantum massive scalar field with arbitrary coupling to a generic spacetime. This expression, which is an expansion in powers of the square of the inverse mass of the quantum field, proportional to coincident limit of the Hadamard-DeWitt coefficient $\left[a_{3}\right]$, is used to obtain the leading approximation for the renormalized quantum stress energy tensor of the quantum field by functional differentiation with respect to the metric. The general expressions obtained are particularized to the case of a background spacetime corresponding to a pointlike global monopole.

The paper is organized as follows. In Section II we present the line element describing a pointlike global monopole, which will be used as a background to quantize the massive scalar field. In Section III we give a brief description of the Schwinger-DeWitt method to construct the quantum effective action, and obtain an analytic expression for this quantity for large mass scalar fields. Section IV is devoted to the construction of the four dimensional renormalized quantum stress tensor $\left<T_{\mu \nu}\right>_{ren}$ for a massive scalar field in a general spacetime in terms of the functional derivatives of the coincident limit of the Hadamard-DeWitt coeficients $a_{3}$. Explicit analytic results for $\left<T_{\mu \nu}\right>_{ren}$ in the spacetime of a pointlike global monopoles are presented and discussed in Section V, whereas Section VI contains our concluding remarks and some perspective about future works on this subject.

\section{The pointlike global monopole spacetime}
The most simple model which gives rise to global monopoles was constructed by Barriola and Vilenkin in \cite{barriola-vilenkin} and starts with the Langrangian density
\begin{equation}
L= {1\over 2} (\partial_\mu \phi^a)( \partial^\mu \phi^a ) - {1 \over
4} \lambda (\phi^a \phi^a - \eta^2)^2\ ,
\end{equation}
where the parameter $\eta$ is of order $10^{16}Gev$ for a typical grand unified theory. From the above Lagrangian density and the Einstein equations, we obtain the spherically symmetric solution:
\begin{equation}
ds^2 = -f(r)dt^2 + f^{-1}(r)dr^2 + r^2(d\theta^2 + \sin^2\theta
d\varphi^2)\ ,
\end{equation}
with $f(r)$, far from the monopole's core is given by
\begin{equation}
f(r) = 1 - 8\pi \eta^2 - 2 M/r\ .
\end{equation}
being $M$ is the mass parameter. If we neglect this mass term we obtain the line element that describes the geometry around a pointlike global monopole, which results in
\begin{equation}
ds^2 = -\alpha^2 dt^2 + dr^2/\alpha^2 + r^2(d\theta^2 +
\sin^2\theta d\varphi^2)\ ,
\label{metric}
\end{equation}
where we define the parameter $\alpha$ according to the expression
\begin{equation}
\alpha^2 = 1 - 8\pi \eta^2.
\label{alfa}
\end{equation}

Re-scaling in the above solution the time and radial variables using $\tau=\alpha t$ and $\rho=\frac{r}{\alpha}$ we arrive to the line element
\begin{equation}
ds^2 = -d\tau^2 + d\rho^2 + \alpha^{2}\rho^2(d\theta^2 +
\sin^2\theta d\varphi^2)\ ,
\label{metric2}
\end{equation}
which shows that this spacetime is characterized by a solid angle deficit, defined as the difference between the solid angle
in the flat spacetime $4\pi$ and the solid angle in the
global monopole spacetime $4\pi\alpha^2$. The parameter $\alpha < 1$ imply a solid angle deficit whereas $\alpha >1$ imply solid angle excess. Taking into account the value of $\eta$ in (\ref{alfa}) we have that field theory predicts a value for $\alpha$ smaller than unity, which imply a solid angle deficit for the pointlike global monopole spacetime.

From the the line element (\ref{metric}) we see that the pointlike global monopole has no Newtonian gravitational
potential, and consequently exerts no gravitational force on the matter around it, apart from the tiny gravitational effect due to the core. However, the geometry around the global monopole has non-vanishing curvature. Then, although the global monopole has no Newtonian
gravitational potential, it gives enormous tidal acceleration proportional to the inverse of the square of the distance from monopole's core, a fact that was considered in reference \cite{HiscockPRL} in a cosmological context to obtain an upper bound on the number
density of them in the Universe, which is at most one
global monopole in the local group of galaxies. In contrast with this result, in \cite{BennetRhie}, the authors show, using numerical simulations, that the real upper boundary is smaller by many orders than that derived by Hiscock in \cite{HiscockPRL}, finding scaling solution which corresponds to a few global monopoles per horizon volume.

\section{The renormalized one-loop effective action}

In this section we construct the one-loop effective action for a massive scalar field with mass $\mu$ and arbitrary coupling to a generic gravitational background with metric tensor $g_{\mu \nu}$, using the Schwinger-DeWitt approach. Details for the results presented in this section can be found in references \cite{frolov-zelnikov,avramidi,matyjasek1,matyjasek,Folacci,owen1,matyjasek3}.

The non-minimally coupled massive scalar field satisfy the Klein-Gordon equation
\begin{equation}
 \left(-\Box + \mu^{2} + \xi R \right) \phi = 0,
\label{kg}
\end{equation}
where $\xi$ is the coupling constant and $R$ is the Ricci scalar.

The one-loop effective action $S^{(1)}$ is related with the Feynman Green's function $G^{F}(x,x')$ of the Klein-Gordon operator in (\ref{kg}) by the expression
\begin{equation}
S^{(1)}=-\frac{i}{2}Tr \ln G^{F}
\label{actiondef}
\end{equation}

In the following we use the Schwinger-DeWitt proper-time formalism which assumes that $G^{F}(x,x')$ is given by
\begin{equation}
G^{F}(x,x') =  \frac{i \Delta^{1/2}}{(4 \pi)^{2}} \int_{0}^{\infty} i ds
\frac{1}{(is)^{2}} \exp\left[-i \mu^{2} s + \frac{i \sigma(x,x')}{2 s} \right]
A(x,x'; is),
\label{grf}
\end{equation}
where
\begin{equation}
 A(x,x'; is) = \sum_{k=0}^{\infty} (is)^k a_{k}(x,x'),
\end{equation}
$s$ is the proper time and the biscalars $a_{k}(x,x')$ are called Hadamard-DeWitt
coefficients. Also $\Delta(x,x')$ is the Van-Vleck-Morette determinant and the biscalar $\sigma(x,x')$
represent one-half of the geodetic distance between the spacetime points $x$ and $x'.$

In four dimensions, the first three terms in (\ref{grf}), respectively proportional to $a_{k}$ with $k=0,\ 1,\ 2$ are divergent. Then, defining the regularized biscalar $A_{reg}(x,x'; is)$ as
\begin{equation}
A_{reg}(x,x'; is) = A(x,x'; is) - \sum_{k=0}^{2}
a_{k}(x,x') (is)^k,
\end{equation}
we can put, in Eq. (\ref{grf}), $A_{reg}(x,x';is)$ instead of $A(x,x'; is)$, which gives finally the regularised four dimensional Green's function $G^{F}_{reg}(x,x')$ as:
\begin{equation}
G^{F}_{reg}(x,x')=\frac{i \Delta^{1/2}}{(4 \pi)^{2}} \int_{0}^{\infty} i ds
\frac{1}{(is)^{2}} \exp\left[-i \mu^{2} s + \frac{i \sigma(x,x')}{2 s} \right]
\sum_{k=3}^{N}
a_{k}(x,x') (is)^k,
\label{gfreg}
\end{equation}

Substituting (\ref{gfreg}) in (\ref{actiondef}) and taking into account the definition of the Trace and the logarithm of an operator given, for example, in reference \cite{birrel} we have for the renormalized one-loop effective action:
\begin{equation}
S_{ren}^{(1)}=\lim_{x'\rightarrow x}\int d^{4}x \sqrt{-g} \ \frac{\Delta^{1/2}}{32\pi^{2}} \int_{0}^{\infty} i ds
\frac{1}{(is)^{3}} \exp\left[-i \mu^{2} s + \frac{i \sigma(x,x')}{2 s} \right]
\sum_{k=3}^{N}
a_{k}(x,x') (is)^k,
\end{equation}

The limiting processes in the above equation gives
\begin{equation}
S_{ren}^{(1)}=\frac{1}{32 \pi^{2}}\int d^{4}x \sqrt{-g} \int_{0}^{\infty} i ds
\frac{1}{(is)^{3}} \exp\left[-i \mu^{2} s \right]\sum_{k=3}^{N}\left[a_{k}\right] (is)^k,
\label{action2}
\end{equation}
where $a_{k} = \lim_{x' \to x} a_{k}(x,x')$ are the coincidence limits of the Hadamard-DeWitt biscalars and the upper sum limit, $N$, gives the order of the Schwinger-DeWitt approximation in $S_{reg}^{(1)}$.
Integrating over $s$ in (\ref{action2}) by making the substitution
$\mu^{2} \to \mu^{2} - i \varepsilon$ ($\varepsilon >0$) \cite{matyjasek3} we obtain for the renormalized one-loop effective action of the massive scalar field the result:
\begin{equation}
S_{ren}^{(1)}=\frac{1}{32 \pi^{2}}\int d^{4}x \sqrt{-g} \sum_{k=3}^{N}\frac{\Gamma\left(k-2\right)}{(\mu^{2})^{k-2}}\left[a_{k}\right],
\end{equation}

which, using the properties of the Gamma function can be written as
\begin{equation}
S_{ren}^{(1)}=\frac{1}{32 \pi^{2}}\int d^{4}x \sqrt{-g} \sum_{k=3}^{N}\frac{\left(k-3\right)!}{(\mu^{2})^{k-2}}\left[a_{k}\right],
\label{main}
\end{equation}

We expect that if the Compton length associated with the field $\lambda_{c},$ is less than the characteristic radius of the curvature of the background geometry, $L$, then a reasonable approximation to $S_{ren}^{(1)}$ is given by the leading term in (\ref{main}), proportional to the inverse of the squared field's mass:

\begin{equation}
S_{ren}^{(1)}=\frac{1}{32 \pi^{2} \mu^{2}}\int d^{4}x \sqrt{-g} \left[a_{3}\right],
\label{mainok}
\end{equation}

The inclusion of higher order terms in the above expansion will be always well motivated, in order to obtain a value for one-loop efective action closely to the exact value of this quantity. However, in the rest of the paper our aim is to take into account only the leading term in (\ref{main}), which imply the calculation of the coincidence limit of the Hadamard-DeWitt coefficient  $\left[a_{3}\right]$.

As we see from (\ref{mainok}), the main task for the calculation of the one-loop effective action in the Schwinger-DeWitt approximation is the determination, up to order $k=3$, of the coincidence limit of the Hadamard-DeWitt biscalar $a_{k}(x,x')$, that satisfy the recurrence equation
\begin{equation}
\sigma^{;i} a_{k;i} + k a_{k} - \Delta^{-1/2}\Box \left( \Delta^{1/2}
a_{k-1}\right) + \xi R a_{k-1} =0,
\label{DeWitt}
\end{equation}
with the boundary condition $a_{0}(x,x')=1.$

The results for this coefficients up to order $k=5$ can be find, for example, in references \cite{DeWitt,avramidi,gilkey,wardel,matyjasek4,owenmonopole1}. Using the transport equation approach of Ottewill and Wardell \cite{wardel}, we can obtain easily general expressions for the coincidence limit of the Hadamard-DeWitt coefficients $a_{3}$, by solving the transport equations given in \cite{wardel} using the software package xAct for Wolfram Mathematica \cite{xact}. The result is:
\begin{eqnarray}
[a_{3}] &=& \frac{1}{15120} (584 R{}_{\alpha }{}_{\beta } R{}^{\alpha }{}_{\gamma } R{}^{\beta }{}^{\gamma }  - 654 R R{}_{\alpha }{}_{\beta } R{}^{\alpha }{}^{\beta } + 99 R^{3} + 456 R{}_{\alpha }{}_{\beta } R{}_{\gamma }{}_{\delta } R{}^{\alpha }{}^{\gamma }{}^{\beta }{}^{\delta } + 72 R R{}_{\alpha }{}_{\beta }{}_{\gamma }{}_{\delta } R{}^{\alpha }{}^{\beta }{}^{\gamma }{}^{\delta } \nonumber \\
&& - 80 R{}_{\alpha }{}_{\beta }{}^{\epsilon }{}^{\rho } R{}^{\alpha }{}^{\beta }{}^{\gamma }{}^{\delta } R{}_{\gamma }{}_{\delta }{}_{\epsilon }{}_{\rho } + 51 R{}_{;\alpha } R{}^{;\alpha } - 12 R{}_{\alpha }{}_{\gamma }{}_{;\beta } R{}^{\alpha }{}^{\beta }{}^{;\gamma } - 6 R{}_{\alpha }{}_{\beta }{}_{;\gamma } R{}^{\alpha }{}^{\beta }{}^{;\gamma } + 27 R{}_{\alpha }{}_{\beta }{}_{\gamma }{}_{\delta }{}_{;\epsilon } R{}^{\alpha }{}^{\beta }{}^{\gamma }{}^{\delta }{}^{;\epsilon } \nonumber \\
&& + 84 R \Box R + 36 R{}_{;\alpha }{}_{\beta } R{}^{\alpha }{}^{\beta } - 24 R{}_{\alpha }{}_{\beta }\Box R{}^{\alpha }{}^{\beta } + 144 R{}_{\alpha }{}_{\beta }{}_{;\gamma }{}_{\delta } R{}^{\alpha }{}^{\gamma }{}^{\beta }{}^{\delta } + 54 \Box \Box R\nonumber \\
&&+ \frac{\xi}{360} (2 R R{}_{\alpha }{}_{\beta } R{}^{\alpha }{}^{\beta } - 5 R^{3} - 2 R R{}_{\alpha }{}_{\beta }{}_{\gamma }{}_{\delta } R{}^{\alpha }{}^{\beta }{}^{\gamma }{}^{\delta } - 12 R{}_{;\alpha } R{}^{;\alpha } - 22 R \Box R - 4 R{}_{;\alpha }{}_{\beta } R{}^{\alpha }{}^{\beta } - 6 \Box \Box R \nonumber \\
&&+\frac{\xi^{2}}{12} (R^{3} + R{}_{;\alpha } R{}^{;\alpha } + 2 R \Box R-\frac{\xi^{3}}{6} R^{3}
\label{a3}
\end{eqnarray}

The above result coincide with those reported in references \cite{wardel,matyjasek4}. As we can see, the coincidence limit of the Hadamard-DeWitt coefficients $\left[a_{3}\right]$ is an extremely complicated local expression constructed from the Riemmann tensor, their covariant derivatives, and contractions. However, the fact that the above result is valid for a generic spacetime, being static, stationary or non-stationary, gives rise to the possibility of use it to obtain relatively simple expressions for the regularized one-loop effective action of the quantum massive scalar field in spacetimes with higher degree of symmetry.

As we can see from the structure of $\left[a_{3}\right]$, it is a local geometric term that depends of the coupling constant $\xi$ and the parameters that describe the geometry of the gravitational background. For the pointlike global monopole spacetime, the only parameter that characterizes the geometry of the manifold is $\alpha$. Then, it is razonable to expect that the one-loop effective action, as well as the regularized quantum stress tensor for the massive scalar field in this background, will be functions of $\xi$, $\alpha$ and the distance $r$ from the monopole core.

Putting (\ref{a3}) into (\ref{mainok}) we can obtain a general expression for the regularized one-loop effective action. However, we can simplify the result using the fact that not all the terms in (\ref{a3}) are independent among themselves. It is possible to show that in four dimensions the following relations holds \cite{Folacci}:
\begin{equation}
\int d^4 x \sqrt{-g} ~\Box \Box R= 0
\end{equation}
\begin{equation}
\int d^4 x \sqrt{-g} ~R_{;\alpha \beta}
R^{\alpha \beta} =   \int d^4 x\sqrt{-g}  \left( \frac{1}{2} \, R\Box R \right)
\end{equation}
\begin{equation}
\int d^4 x \sqrt{-g} ~R_{\alpha \beta ; \gamma \delta}R^{\alpha \gamma \beta \delta} =\int d^4 x  \sqrt{-g} \left( -\frac{1}{4} \,  R\Box
R + R_{\alpha \beta} \Box R^{\alpha \beta}  - R_{\alpha \beta} R^{\alpha}_{\gamma}R^{\beta \gamma} + R_{\alpha \beta}R_{\gamma \delta}R^{\alpha \gamma \beta \delta} \right)
\end{equation}
\begin{equation}
 \int d^4 x\sqrt{-g} ~R_{;\alpha}R^{;\alpha} =
\int d^4 x \sqrt{-g} \left(-R\Box R\right)
\end{equation}
\begin{equation}
\int d^4 x\sqrt{-g} ~R_{\alpha \beta;\gamma}
R^{\alpha \beta;\gamma} = \int d^4 x\sqrt{-g}  \left(-R_{\alpha \beta} \Box R^{\alpha \beta}\right)
\end{equation}
\begin{equation}
\int d^4 x\sqrt{-g} ~R_{\alpha \beta;\gamma} R^{\alpha \gamma;\beta} = \int d^4 x \sqrt{-g} \left( -\frac{1}{4} R\Box R
-R_{\alpha \beta} R^{p}_{\alpha
\gamma}R^{\beta \gamma} + R_{\alpha \beta}R_{\gamma \delta}R^{\alpha \gamma \beta \delta} \right)
\end{equation}
\begin{eqnarray}
\int d^4 x \sqrt{-g} ~R_{\alpha \beta \gamma \delta;\epsilon} R^{\alpha \beta \gamma \delta ;\epsilon}= \int d^4 x \sqrt{-g} \left(  R\Box R -4 R_{\alpha \beta} \Box
R^{\alpha \beta} + 4  R_{\alpha \beta} R^{\alpha}_{\beta \gamma}R^{\beta \gamma} - 4
R_{\alpha \beta}R_{\gamma \delta}R^{\alpha \gamma \beta \delta} \right.\nonumber \\
\left. -2 R_{\alpha \beta}R^\alpha_{\alpha \gamma \delta \epsilon} R^{\beta \gamma \delta \epsilon } +
R_{\alpha \beta \gamma \delta}R^{\alpha \beta \mu \nu} R^{\gamma \delta}_{\gamma \delta \mu \nu} + 4 R_{\alpha \gamma \beta \delta} R^{\alpha
\mu \beta}_{\alpha \mu \beta \nu} R^{\gamma \mu \delta \nu} \right).
\nonumber \\
\end{eqnarray}

Using the above relations we can show that the one-loop effective action only includes ten geometric terms. The final result is
\begin{eqnarray} \label{efactionok}
& & S^{(1)}_{\mathrm{ren}}= \frac{1}{192 \pi^2 \mu^2}\int d^4
x\sqrt{-g} ~\left[\left(\frac{\xi^2}{2}- \frac{\xi}{5}+\frac{1}{56}\right)
  \, R\Box R  +\frac{1}{140}\,  R_{\alpha \beta} \Box R^{\alpha \beta}  - \left(\xi-\frac{1}{6}\right)^3 \, R^3  \right. \nonumber \\
& & \qquad \left.   + \frac{1}{30}\, \left(\xi-\frac{1}{6}\right)
\,  RR_{\alpha \beta} R^{\alpha \beta}  - \frac{8}{945} \, R_{\alpha \beta} R^{\alpha}_{\phantom{\alpha} \gamma}R^{\beta \gamma} +\frac{2}{315}
\, R_{\alpha \beta}R_{\gamma \delta}R^{\alpha \gamma \beta \delta}+ \frac{1}{1260} \, R_{\alpha \beta}R^\alpha_{\phantom{\alpha} \gamma \delta \epsilon} R^{\beta \gamma \delta \epsilon} \right. \nonumber \\
& &  \qquad \left.
      - \frac{1}{30}\, \left(\xi-\frac{1}{6}\right) \, RR_{\alpha \beta \gamma \delta} R^{\alpha \beta \gamma \delta}+ \frac{17}{7560} \, R_{\alpha \beta \gamma \delta}R^{\alpha \beta \sigma \rho}
R^{\gamma \delta}_{\phantom{\gamma \delta} \sigma \rho} - \frac{1}{270}\, R_{\alpha \gamma \beta \delta} R^{\alpha \phantom{\sigma}
\beta}_{\phantom{\alpha} \sigma \phantom{\beta} \rho} R^{\gamma \sigma \delta \rho} \right].
\end{eqnarray}

\section{Schwinger-DeWitt approximation for $\left<T_{\mu}^{\nu}\right>_{ren}$}

The renormalized quantum stress energy tensor for the massive scalar field nonminimally coupled to a generic spacetime background can be determined from (\ref{efactionok}) by functional differentiation with respect to the metric tensor:
\begin{equation}\label{SETdef}
 \langle  ~T^{\mu\nu
} ~ \rangle_{\mathrm{ren}} =\frac{2}{ \sqrt{-g}} \frac{\delta
S^{(1)}_{\mathrm{ren}}} {\delta
g_{\mu\nu }}
\end{equation}

Due to the identities and relations satisfied by the Riemmann tensor, its contractions and covariant derivatives, there are not unique results for $ \langle  ~T^{\mu\nu} ~ \rangle_{\mathrm{ren}} $ \cite{matyjasek1,matyjasek,owen1,owen2,Folacci}. However, all the obtained expressions for this quantity must give the same results when applied to specific spacetime backgrounds. Also, all the results must have covariant divergence equal to zero, which is a fundamental property of the stress energy tensor. 

In the following we use the basis proposed in the beautiful paper of Decaninis and Follaci \cite{Folacci}, which allow us to obtain an irreducible expression for the renormalized stress energy tensor for the quantum scalar field given by :
\begin{eqnarray}\label{SET}
&&  \langle  ~T_{\mu\nu
} ~  \rangle_{\mathrm{ren}}    = \frac{1}{96 \pi^2 \mu^2}\left[ \left[\xi^2-\frac{2\xi}{5}+\frac{3}{70}\right]  \, (\Box R)_{;\mu\nu}
+ \frac{1}{10}\left(\xi-\frac{1}{6}\right)  \, R \Box R_{\mu \nu} + \frac{1}{15}\left(\xi-\frac{1}{7}\right)
\,R_{;\alpha (\mu} R^{\alpha}_{\phantom{\alpha}\nu)} \right. \nonumber\\
&& \qquad \qquad \left. -\frac{1}{140} \,  \Box \Box R_{\mu \nu} -6\,\left(\xi-\frac{1}{6}\right)\left[\xi^2-\frac{\xi}{3}+\frac{1}{30}\right] \,  R R_{;\mu \nu}-\left(\xi-\frac{1}{6}\right)\left(\xi-\frac{1}{5}\right)\, (\Box R) R_{\mu \nu} \right.\nonumber \\
&& \qquad \qquad \left. + \frac{1}{42} \, R_{\alpha (\mu}
\Box
R^\alpha_{\phantom{\alpha} \nu)}  + \frac{1}{15}\left(\xi-\frac{2}{7}\right) \,R^{\alpha \beta} R_{\alpha \beta;(\mu \nu)}
 + \frac{2}{105} \, R^{\alpha \beta} R_{\alpha (\mu ; \nu)\beta}-\frac{1}{70} \, R^{\alpha \beta} R_{\mu\nu; \alpha \beta} \right. \nonumber\\
&& \qquad \qquad \left.
  + \frac{2}{15}\left(\xi-\frac{3}{14}\right) \,R^{;\alpha \beta}R_{\alpha \mu \beta \nu}  -\frac{1}{105} \, (\Box R^{\alpha \beta})R_{\alpha \mu \beta \nu}+
 \frac{13}{105} \,R^{\alpha \beta;\gamma}_{\phantom{\alpha \beta;\gamma} (\mu} R_{|\gamma \beta \alpha|
\nu)}   \right. \nonumber \\
&& \qquad \qquad \left. + \frac{2}{35}  \,R^{\alpha \phantom{(\mu}; \beta \gamma}_{\phantom{\alpha } (\mu} R_{|\alpha \beta \gamma|
\nu)} -\frac{1}{15}\left(\xi-\frac{3}{14}\right) \,R^{\alpha \beta \gamma \delta} R_{\alpha \beta \gamma \delta ; (\mu \nu)
}  -6\left(\xi-\frac{1}{4}\right)\left(\xi-\frac{1}{6}\right)^2 \, R_{;\mu} R_{;\nu} \right. \nonumber \\
&& \qquad \qquad \left.  -\frac{1}{5}\left(\xi-\frac{3}{14}\right)  \,
R_{;\alpha } R^\alpha_{\phantom{\alpha} (\mu;\nu)}+ \frac{1}{5}\left(\xi-\frac{17}{84}\right)  \,R_{;\alpha }R_{\mu\nu}^{\phantom{\mu\nu}; \alpha}
+ \frac{1}{15}\left(\xi-\frac{1}{4}\right)  \, R^{\alpha \beta}_{\phantom{\alpha \beta};\mu}R_{\alpha \beta;\nu}\right. \nonumber \\
&& \qquad \qquad \left. -\frac{1}{210} \,
R^\alpha_{\phantom{\alpha} \mu;\beta} R_{\alpha \nu}^{\phantom{\alpha \nu};\beta} +
 \frac{1}{42} \,R^\alpha_{\phantom{\alpha} \mu;\beta} R^\beta_{\phantom{\beta} \nu;\alpha} -\frac{1}{105}\,
R^{\alpha \beta;\gamma} R_{\gamma \beta \alpha (\mu;\nu) } -\frac{1}{70} \, R^{\alpha \beta;\gamma} R_{\alpha \mu \beta \nu;\gamma} \right. \nonumber \\
&& \qquad \qquad \left. -\frac{1}{15}\left(\xi-\frac{13}{56}\right) \,R^{\alpha \beta \gamma \delta}_{\phantom{\alpha \beta \gamma \delta};\mu} R_{\alpha \beta \gamma \delta ; \nu }-\frac{1}{70} \, R^{\alpha \beta \gamma}_{\phantom{\alpha \beta \gamma}\mu;\delta}R_{\alpha \beta \gamma \nu}^{\phantom{\alpha \beta \gamma
\nu};\delta}   + 3\left(\xi-\frac{1}{6}\right)^3  \, R^2 R_{\mu\nu}\right. \nonumber \\
&& \qquad \qquad \left. -\frac{2}{15}\left(\xi-\frac{1}{6}\right) \, R R_{\alpha \mu} R^\alpha_{\phantom{\alpha}\nu}
-\frac{1}{30}\left(\xi-\frac{1}{6}\right) \, R^{\alpha \beta}R_{\alpha \beta}R_{\mu\nu} -\frac{2}{315} \, R^{\alpha \beta}R_{\alpha \mu}R_{\beta \nu}\right. \nonumber \\
&& \qquad \qquad \left. + \frac{1}{315} \,R^{\alpha \gamma}R^\beta_{\phantom{\beta} r}R_{\alpha \mu \beta \nu} + \frac{1}{315} \, R^{\alpha \beta}R^\gamma_{\phantom{\gamma} (\mu}R_{ |\gamma \beta \alpha|  \nu) }
+ \frac{1}{15}\left(\xi-\frac{1}{6}\right) \, R R^{\alpha \beta \gamma}_{\phantom{\alpha \beta \gamma}\mu }R_{\alpha \beta \gamma \nu} \right. \nonumber \\
&& \qquad \qquad\left.
 + \frac{1}{30}\left(\xi-\frac{1}{6}\right) \,R_{\mu\nu}R^{\alpha \beta \gamma \delta} R_{\alpha \beta \gamma \delta }
 -\frac{4}{315}\, R^\alpha_{\phantom{\alpha}  (\mu}R^{\beta \gamma \delta}_{\phantom{\beta \gamma \delta} |\alpha|
}R_{ |\beta \gamma \delta| \nu) } -\frac{2}{315} \, R^{\alpha \beta}R^{\gamma \delta}_{\phantom{\gamma \delta} p\mu}R_{\gamma \delta \beta \nu}\right. \nonumber \\
&& \qquad \qquad \left.+ \frac{4}{315} \, R_{\alpha \beta}R^{\alpha \gamma \beta \delta}R_{\gamma \mu \delta \nu} -\frac{1}{315} \, R_{\alpha \beta}R^{\alpha \gamma \delta}_{\phantom{\alpha \gamma \delta}\mu}R^\beta_{\phantom{\beta} \gamma \delta \nu}+ \frac{2}{315} \,R^{\alpha \beta \gamma \delta}R_{\alpha \beta \sigma \mu }R_{\gamma \delta \phantom{\sigma}
\nu}^{\phantom{\gamma \delta} \sigma}\right. \nonumber \\
&& \qquad \qquad \left.
 + \frac{4}{63} \, R^{\alpha \gamma \beta \delta}R^\sigma_{\phantom{\sigma}
\alpha \beta \mu}R_{\sigma \gamma \delta \nu}  -\frac{2}{315} \,  R^{\alpha \beta \gamma}_{\phantom{\alpha \beta \gamma} \delta } R_{\alpha \beta \gamma \sigma}R^{\delta
\phantom{\mu} \sigma}_{\phantom{\delta} \mu \phantom{\sigma} \nu} + \frac{1}{15}\left(\xi-\frac{1}{6}\right) \, R R^{\alpha \beta}R_{\alpha \mu \beta \nu}\right. \nonumber \\
&&  \qquad \qquad \left.
+ g_{\mu\nu} \left[  \left(-\xi^2+ \frac{2\xi}{5}-\frac{11}{280}\right) \, \Box \Box R + 6\left(\xi-\frac{1}{6}\right)\left[\xi^2- \frac{\xi}{3}+\frac{1}{40}\right]  \, R\Box R \right. \right. \nonumber\\
&& \qquad \qquad \left. \left.
-\frac{1}{30}\left(\xi-\frac{3}{14}\right)  \, R_{;\alpha \beta}
R^{\alpha \beta} -\frac{1}{15}\left(\xi-\frac{5}{28}\right)  \,  R_{\alpha \beta} \Box R^{\alpha \beta}+ \frac{4}{15}\left(\xi-\frac{1}{7}\right)  \, R_{\alpha \beta ; \gamma \delta}R^{\alpha \gamma \beta \delta} \right. \right.\nonumber \\
&& \qquad \qquad \left. \left. + 6\left(\xi^3-\frac{13}{24}\xi^2+\frac{17}{180}\xi-\frac{53}{10080}\right)  \,R_{;\alpha}R^{;\alpha}-\frac{1}{15}\left(\xi-\frac{13}{56}\right)  \, R_{\alpha \beta;\gamma} R^{\alpha \beta;\gamma} \right.\right.  \nonumber \\
&& \qquad  \qquad \left. \left. -\frac{1}{420} \, R_{\alpha \beta;\gamma}
R^{\alpha \gamma;\beta} + \frac{1}{15}\left(\xi-\frac{19}{22}\right)  \,R_{\alpha \beta \gamma \delta;\sigma} R^{\alpha \beta \gamma \delta;\sigma}
     -\frac{1}{2}\left(\xi-\frac{1}{6}\right)^{3}  \, R^3  \right.\right. \nonumber \\
&& \qquad \qquad \left.\left. + \frac{1}{60}\left(\xi-\frac{1}{6}\right) \, RR_{\alpha \beta} R^{\alpha \beta} + \frac{1}{1890} \,R_{\alpha \beta}
R^{\alpha}_{\phantom{\alpha} r}R^{\beta \gamma}  + -\frac{1}{630} \, R_{\alpha \beta}R_{\gamma \delta}R^{\alpha \gamma \beta \delta}  \right. \right. \nonumber \\
&& \qquad  \qquad \left. \left.-\frac{1}{60}\left(\xi-\frac{1}{6}\right) \,
RR_{\alpha \beta \gamma \delta} R^{\alpha \beta \gamma \delta}
 + \frac{2}{15}\left(\xi-\frac{1}{6}\right) \,R_{\alpha \beta}R^\alpha_{\phantom{\alpha} \gamma \delta \sigma} R^{\beta \gamma \delta \sigma }\right. \right. \nonumber \\
&& \qquad \qquad \left. \left. -\frac{1}{15}\left(\xi-\frac{47}{252}\right) \, R_{\alpha \beta \gamma \delta}R^{\alpha \beta \sigma \rho}
R^{\gamma \delta}_{\phantom{\gamma \delta} \sigma \rho} -\frac{4}{15}\left(\xi-\frac{41}{252}\right) \, R_{\alpha \gamma \beta \delta} R^{\alpha \phantom{\sigma}
\beta}_{\phantom{\alpha} \sigma \phantom{\beta} \rho} R^{\gamma \sigma \delta \rho}\right ]\right].
\end{eqnarray}

The above result is a rather complex expression for the renormalized stress energy tensor for a large mass scalar field with arbitrary coupling to gravity in the Schwinger-DeWitt approximation, that is valid for any spacetime \cite{belokogne}. As we can see the information of the massive scalar field is included in the coefficients accompanying each local geometric term constructed from the Riemmann tensor, its covariant derivatives and contractions.

The fact that we have an analytic result for $\left<T_{\mu \nu}\right>_{ren}$ valid for a generic gravitational background is very important, because opens the possibility to study the influence of the quantization of massive scalar field upon the background spacetime, the so called backreaction problem. Using the above stress energy tensor as a source in the semiclassical Einstein's equations we can, in principle, find the quantum corrections to the background metric perturbatively.

In the rest of the paper we will apply the above obtained formula to the study of the renormalized stress energy tensor for the quantized nonminimally coupled massive scalar field in the spacetime of a pointlike global monopole. As we will show, simple results can be obtained in this case.

\section{Renormalized quantum stress energy tensor for massive scalar field in pointlike global monopole spacetime}

Using (\ref{metric}) in (\ref{SET}) we obtain, for the temporal component of the renormalized stress energy tensor for the massive scalar field with arbitrary coupling parameter $\xi$ the very simple result
\begin{equation}
\left<T_{t}^{t}\right>_{ren}=\frac{\left({\alpha}^{2}-1\right)}{10080 \, \pi^{2}\mu^{2}r^{6}}\sum_{k=0}^{3}B_{k}(\alpha)\xi^{k}
\end{equation}
where
\begin{equation}
B_{0}(\alpha)=101\,{
\alpha}^{2}\left(1+\,{\alpha}^{2}\right)-4
\end{equation}
\begin{equation}
B_{1}=-504\,{\alpha}^{4}-1554\,{\alpha}^{2}+42
\end{equation}
\begin{equation}
B_{2}=-3150\,{\alpha}^{4}+8400\,{\alpha}^{2}-210
\end{equation}
and
\begin{equation}
B_{3}=15540\,{\alpha}^{4}-15960\,{\alpha}^{2}+420
\end{equation}

\begin{figure} 
\centering
\includegraphics[width=10cm]{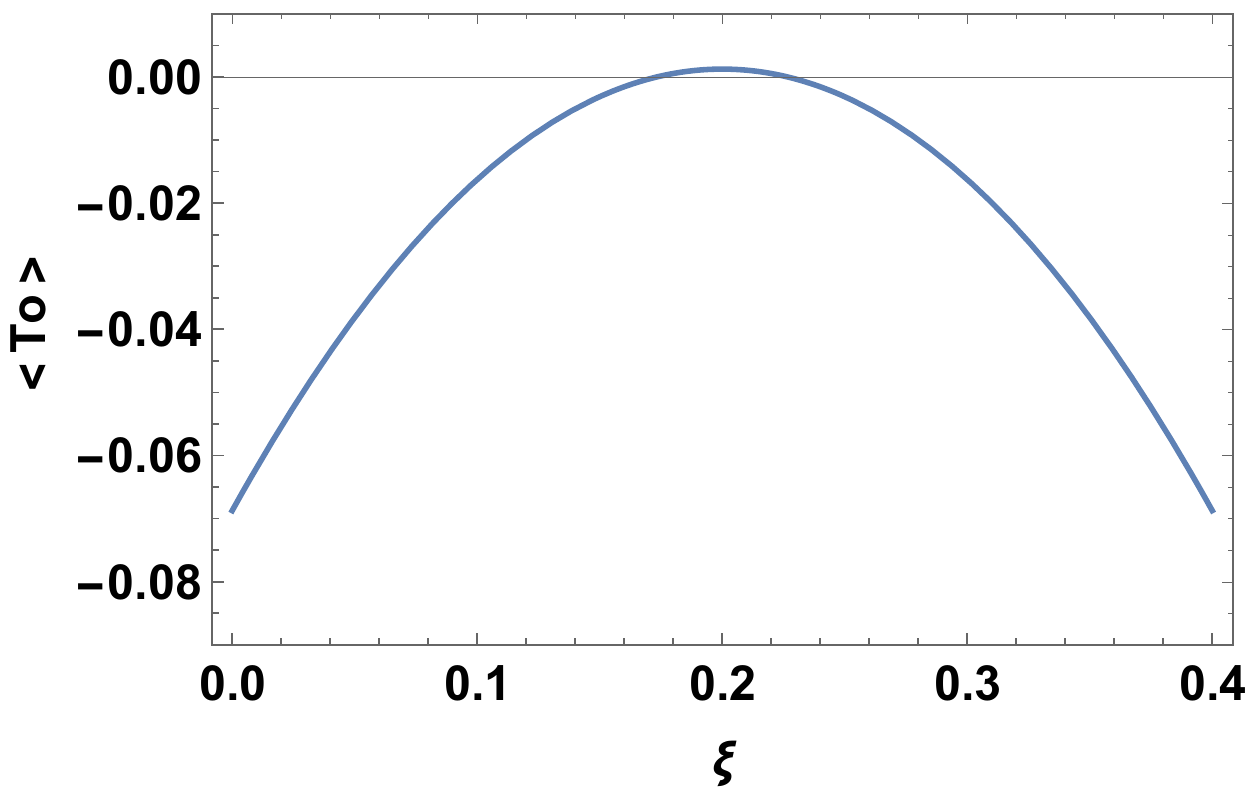}
\caption{ \textit{Dependance on the coupling constant $\xi$ of the re-scaled time component of the renormalized stress energy tensor $\left<T_{0}\right>=96\pi^{2}\mu^{2}\left <T_{t}^{t}\right >_{ren}$ for massive scalar field in the pointlike global monopole spacetime. The values of the parameters used in the calculations are $r=\frac{1}{3}$, and $1 - \alpha^2 =10^{-5}$}.}
\label{f1}
\end{figure}

In Figure \ref{f1} we show the dependance on the coupling constant $\xi$ of the re-scaled time component of the renormalized stress energy tensor $\left<T_{0}\right>=96\pi^{2}\mu^{2}\left <T_{t}^{t}\right >_{ren}$ for massive scalar field in the pointlike global monopole spacetime with parameter $1 - \alpha^2 =10^{-5}$ at a fixed distance from monopole's core.

\begin{figure} 
\centering
\includegraphics[width=10cm]{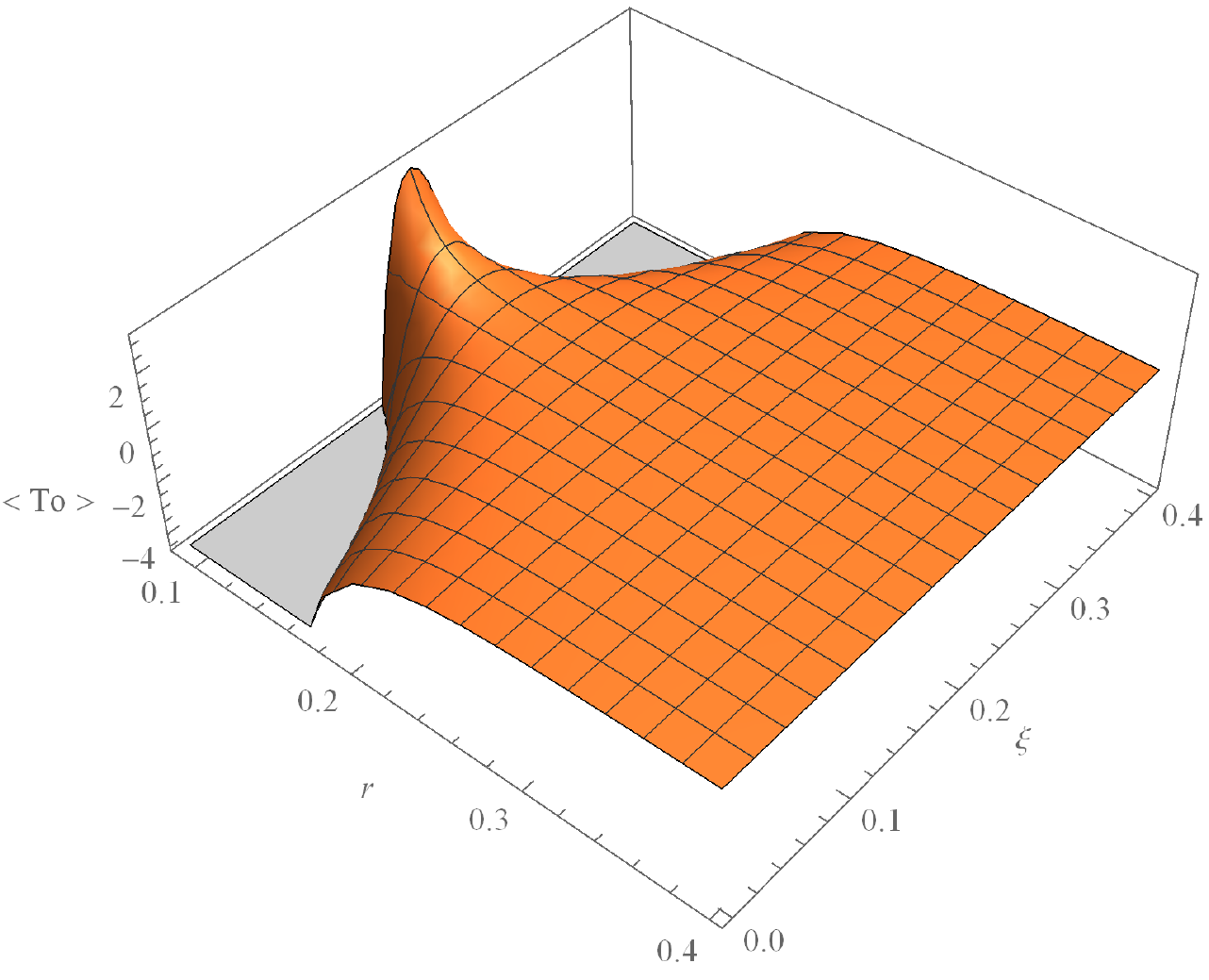}
\caption{ \textit{Dependance on the coupling constant $\xi$ and distance from monopole's core $r$ of the re-scaled time component of the renormalized stress energy tensor $\left<T_{0}\right>=96\pi^{2}\mu^{2}\left <T_{t}^{t}\right >_{ren}$ for massive scalar field in the pointlike global monopole spacetime. The value of the parameter used in the calculations is $1 - \alpha^2 =10^{-5}$}.}
\label{f2}
\end{figure}
The first thing that we can observe is that $\left<T_{0}\right>$ increases with the increase of the coupling constant until it reach its maximum value at $\xi=0.2$, becoming positive for values of the coupling constant between $0.17\leq\xi\leq0.23$. For values of the coupling constant outside this interval the stress energy tensor is negative, decreasing its value for $\xi\geq 0.23$.
\begin{figure}[t]
\scalebox{0.69}{\includegraphics{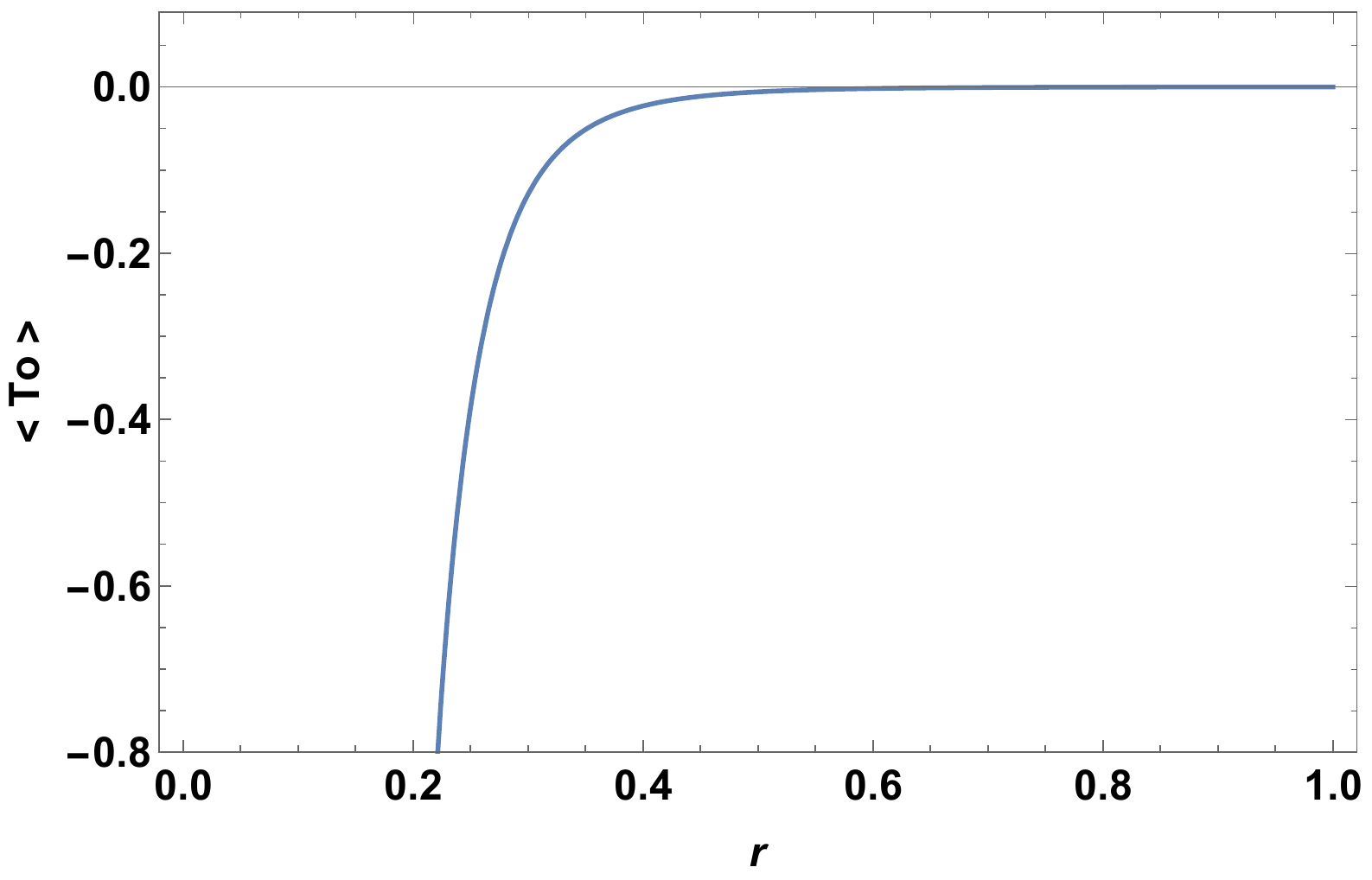}}
\vspace{0.15cm}
\scalebox{0.67}{\includegraphics{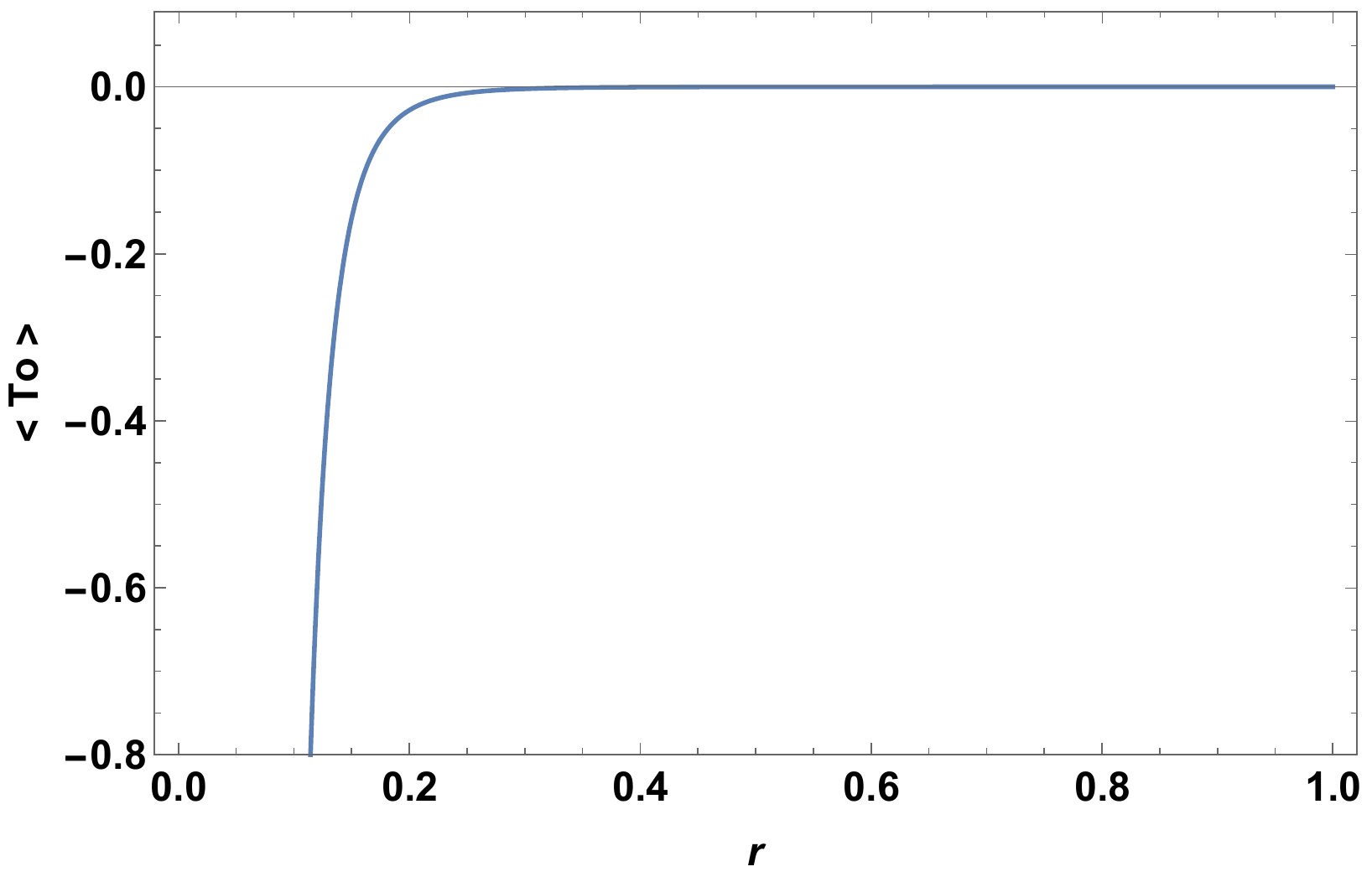}}
\caption{\textit{Dependance on the distance from monopole's core $r$ of the re-scaled time component of the renormalized stress energy tensor $\left<T_{0}\right>=96\pi^{2}\mu^{2}\left <T_{t}^{t}\right >_{ren}$ for massive scalar field in the pointlike global monopole spacetime. The values of the parameters used in the calculations are $1 - \alpha^2 =10^{-5}$ and $\xi=0 (top \ plot)$ and $\xi=\frac{1}{6} (bottom \ plot)$}.} \label{f3}
\end{figure}

The above behavior is similar for all values of the distance $r$ from monopole's core, as we can see in Figure \ref{f2}, where we show the dependance of $\left<T_{0}\right>$ as a function of $\xi$ and $r$. As we can see from the graph, for values of $0.17\leq\xi\leq0.23$, the time component of the stress energy tensor for the massive scalar field is positive, and tends to zero at large distances from  monopole's core. This behaviour excludes the physical values associated with minimal coupling $\xi=0$ and conformal coupling $\xi=\frac{1}{6}$.

For values of the coupling constant outside the above mentioned interval, the time component of the renormalized stress energy tensor has negative values, again tending to zero as $r\rightarrow \infty$. The physical values corresponding to minimal and conformal coupling to gravity are included in this case, as we can see in Figure \ref{f3}, when we plot the dependance of the re-scaled time component of the renormalized stress energy tensor as a function of $r$ for this values of the coupling constant.

\begin{figure} 
\centering
\includegraphics[width=10cm]{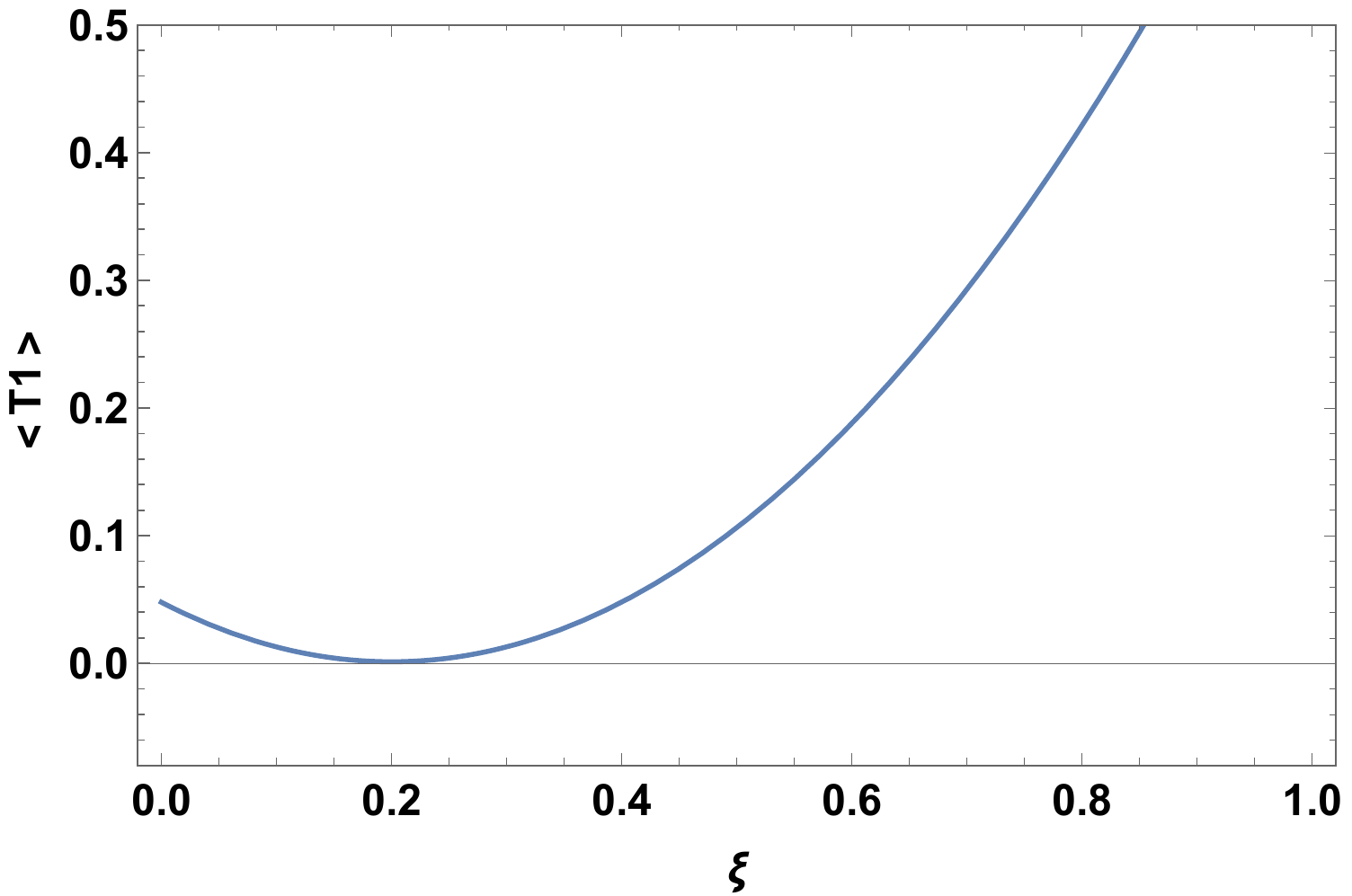}
\caption{\textit{ Dependance on the coupling constant $\xi$ of the re-scaled radial component of the renormalized stress energy tensor $\left<T_{1}\right>=96\pi^{2}\mu^{2}\left <T_{r}^{r}\right >_{ren}$ for massive scalar field in the pointlike global monopole spacetime. The values of the parameters used in the calculations are $r=\frac{1}{3}$, and $1 - \alpha^2 =10^{-5}$}.}
\label{f4}
\end{figure}

\begin{figure} 
\centering
\includegraphics[width=10cm]{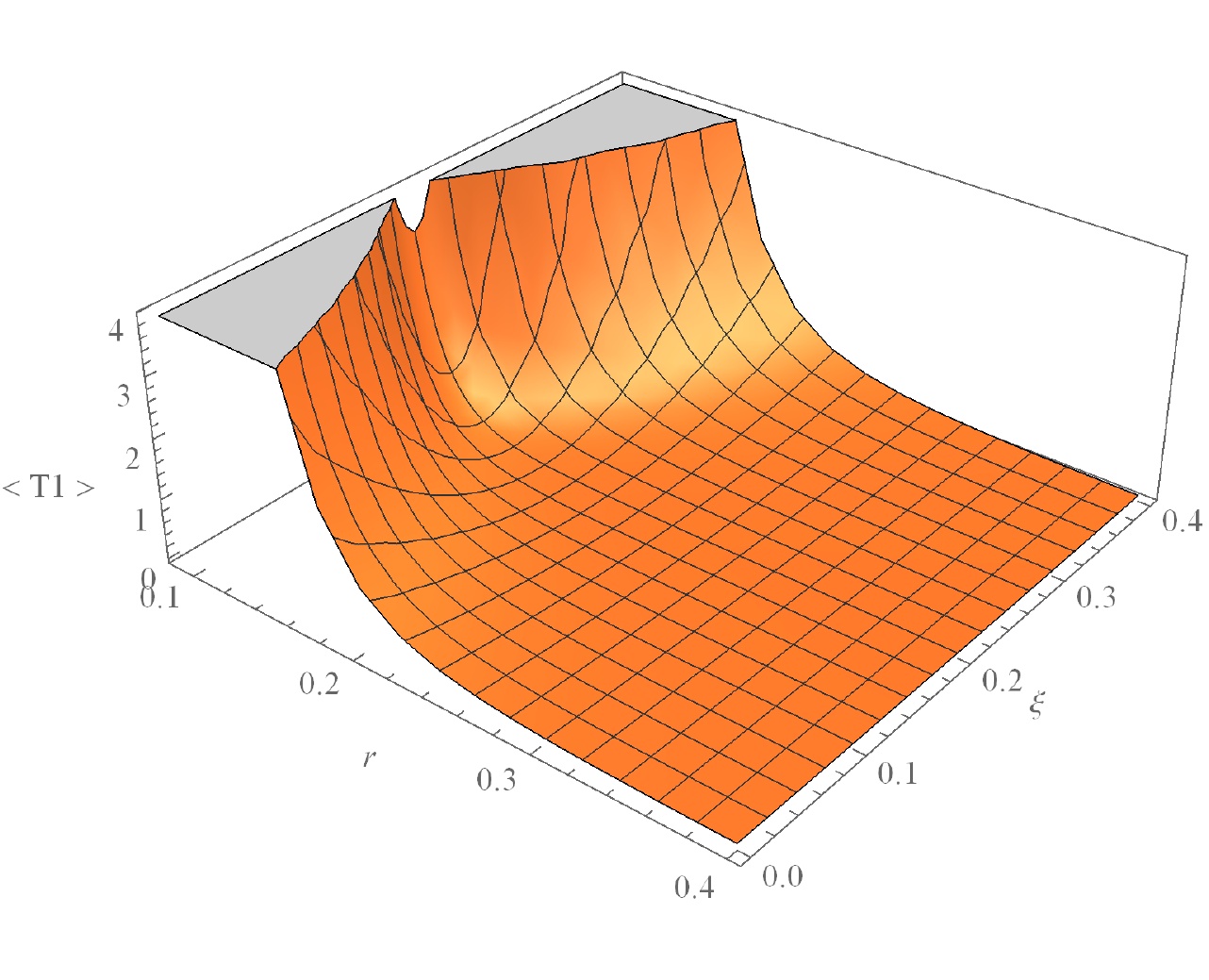}
\caption{ \textit{Dependance on the coupling constant $\xi$ and the distance $r$ from monopole's core of the re-scaled radial component of the renormalized stress energy tensor $\left<T_{1}\right>=96\pi^{2}\mu^{2}\left <T_{r}^{r}\right >_{ren}$ for massive scalar field in the pointlike global monopole spacetime. The value of the parameter used in the calculations is $1 - \alpha^2 =10^{-5}$}.}
\label{f5}
\end{figure}

\begin{figure}[t]
\scalebox{0.69}{\includegraphics{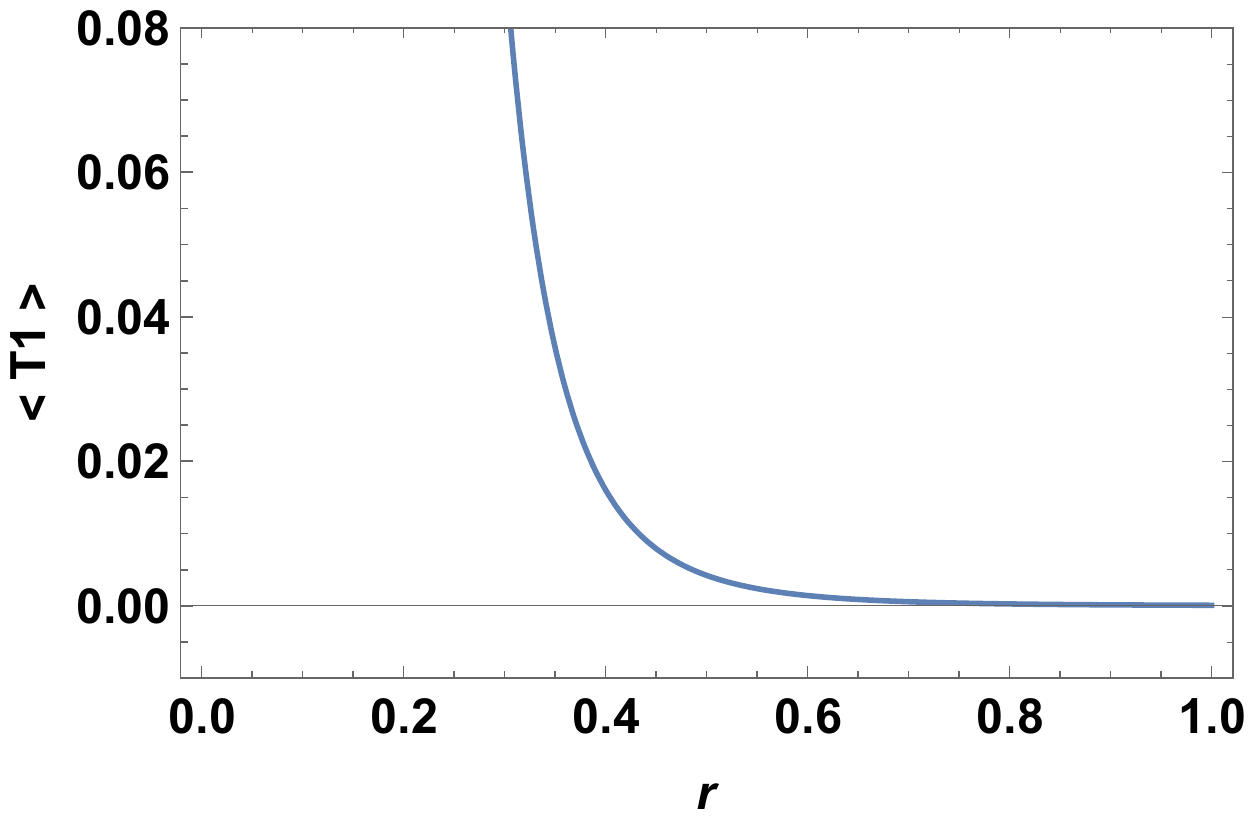}}
\vspace{0.15cm}
\scalebox{0.67}{\includegraphics{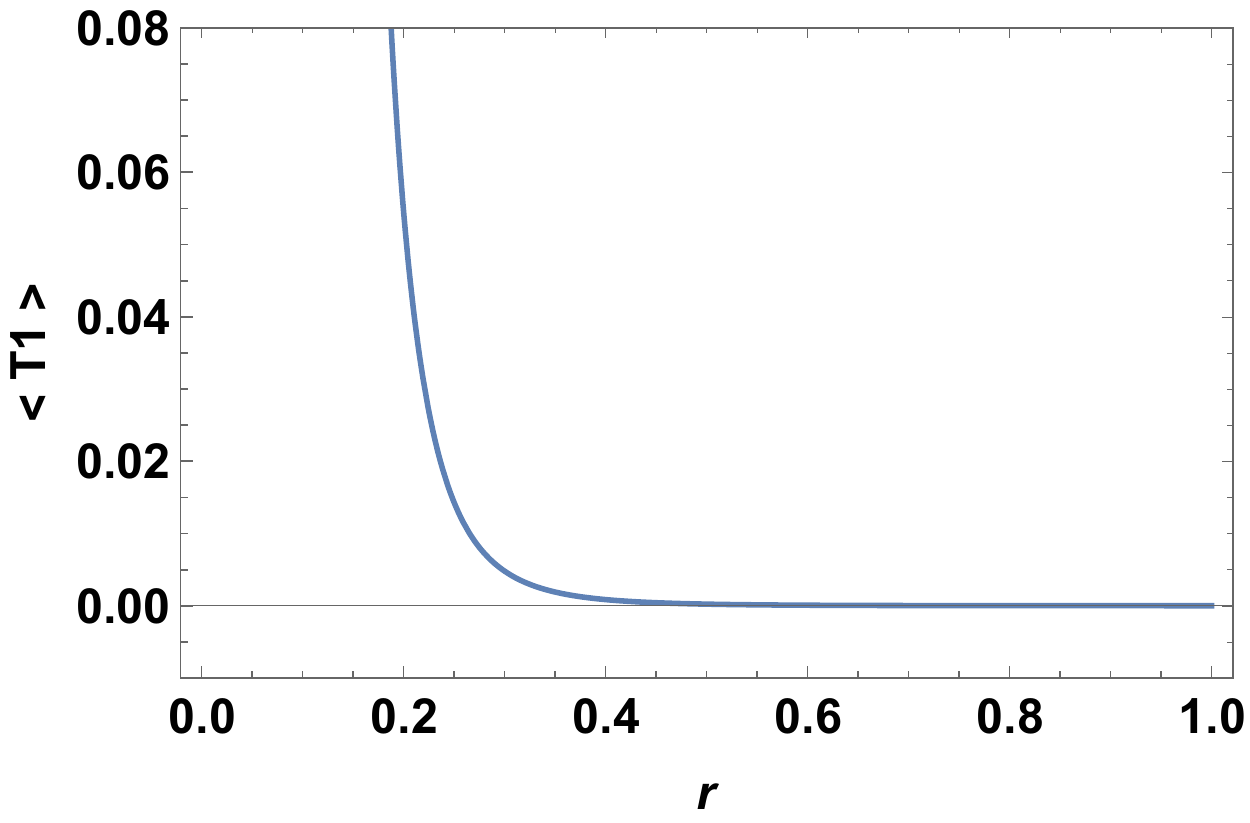}}
\caption{\textit{Dependance on the distance from monopole's core $r$ of the re-scaled radial component of the renormalized stress energy tensor $\left<T_{1}\right>=96\pi^{2}\mu^{2}\left <T_{r}^{r}\right >_{ren}$ for massive scalar field in the pointlike global monopole spacetime. The values of the parameters used in the calculations are $1 - \alpha^2 =10^{-5}$ and $\xi=0 (top \ plot)$ and $\xi=\frac{1}{6} (bottom \ plot)$}.} \label{f6}
\end{figure}

\begin{figure} 
\centering
\includegraphics[width=10cm]{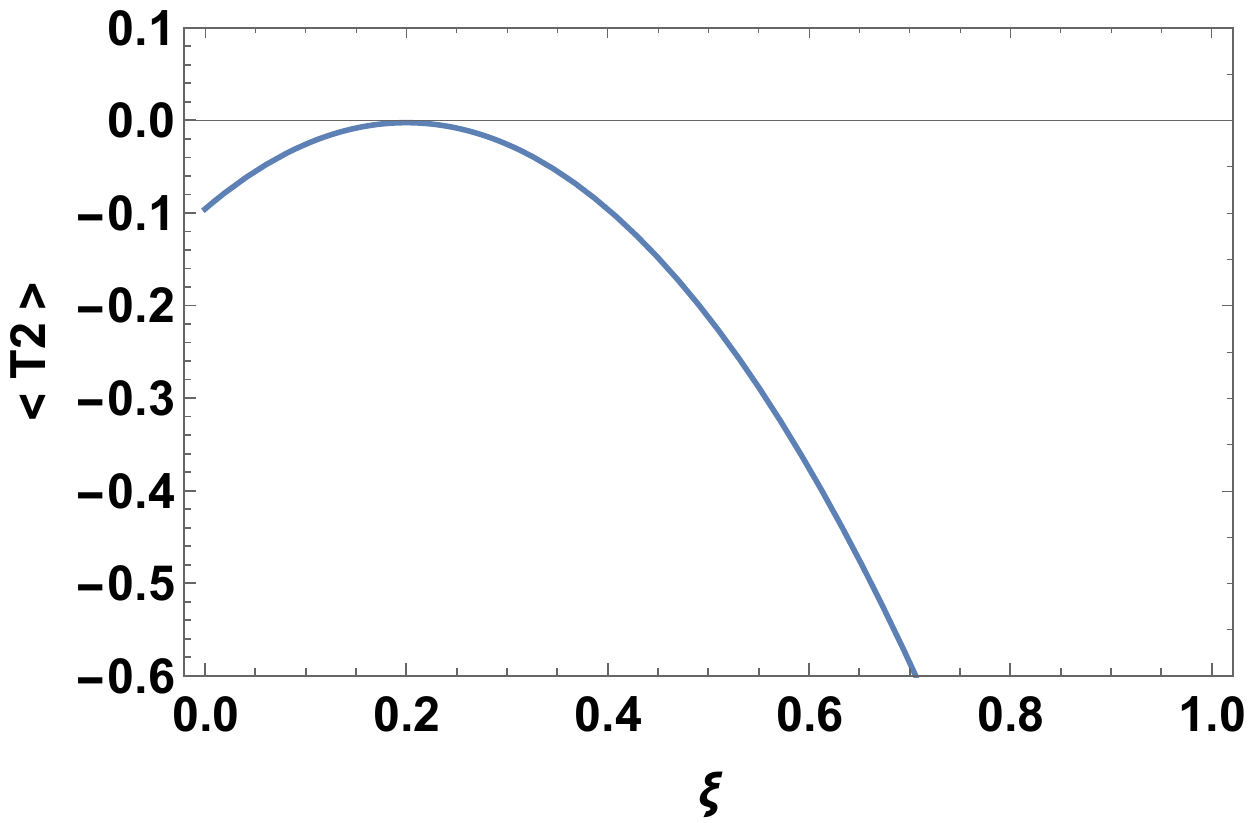}
\caption{ \textit{Dependance on the coupling constant $\xi$ of the re-scaled angular component of the renormalized stress energy tensor $\left<T_{2}\right>=96\pi^{2}\mu^{2}\left <T_{\theta}^{\theta}\right >_{ren}=96\pi^{2}\mu^{2}\left <T_{\phi}^{\phi}\right >_{ren}$ for massive scalar field in the pointlike global monopole spacetime. The values of the parameters used in the calculations are $r=\frac{1}{3}$, and $1 - \alpha^2 =10^{-5}$}.}
\label{f7}
\end{figure}
\begin{figure} 
\centering
\includegraphics[width=10cm]{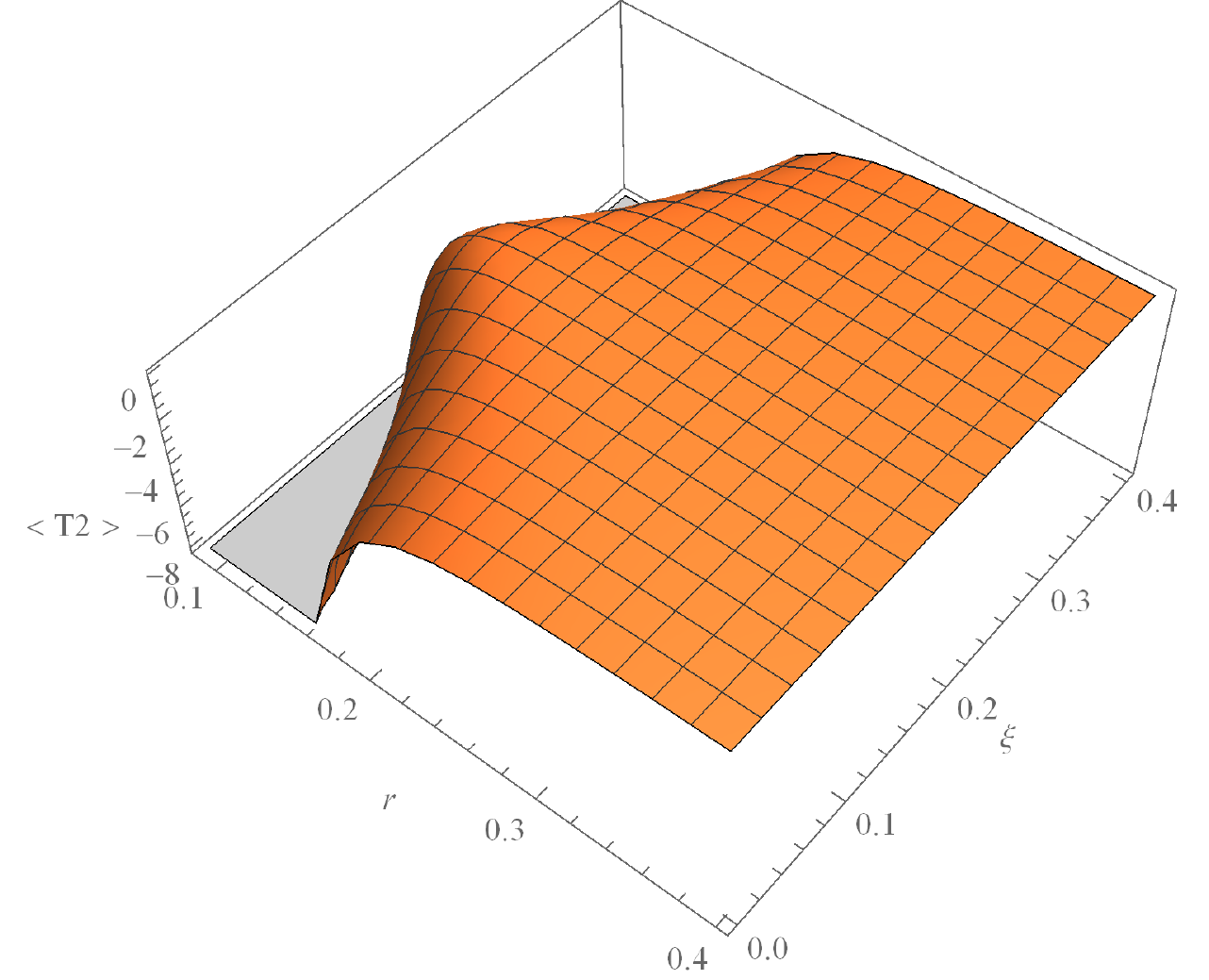}
\caption{ \textit{Dependance on the coupling constant $\xi$ and the distance $r$ from monopole's core of the re-scaled angular component of the renormalized stress energy tensor $\left<T_{2}\right>=96\pi^{2}\mu^{2}\left <T_{\theta}^{\theta}\right >_{ren}=96\pi^{2}\mu^{2}\left <T_{\phi}^{\phi}\right >_{ren}$ for massive scalar field in the pointlike global monopole spacetime. The value of the parameter used in the calculations is $1 - \alpha^2 =10^{-5}$}.}
\label{f8}
\end{figure}

\begin{figure}[t]
\scalebox{0.69}{\includegraphics{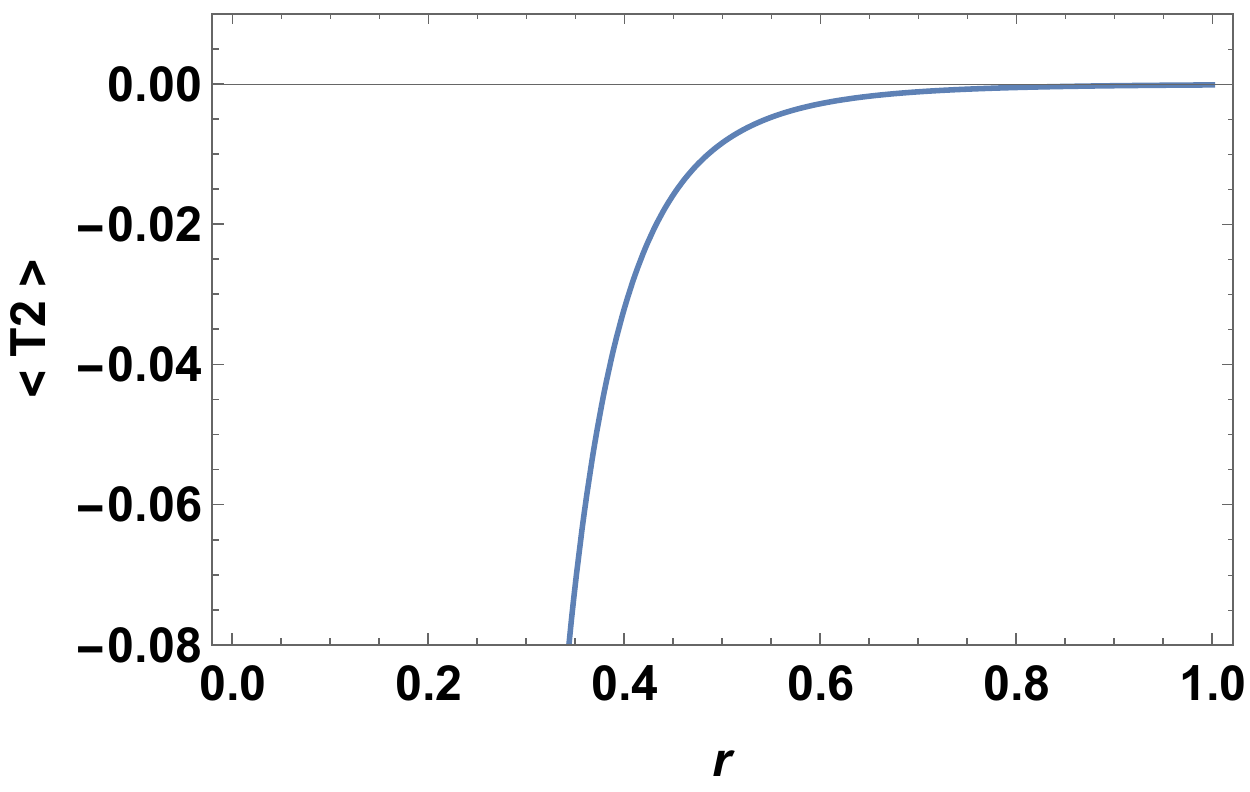}}
\vspace{0.15cm}
\scalebox{0.67}{\includegraphics{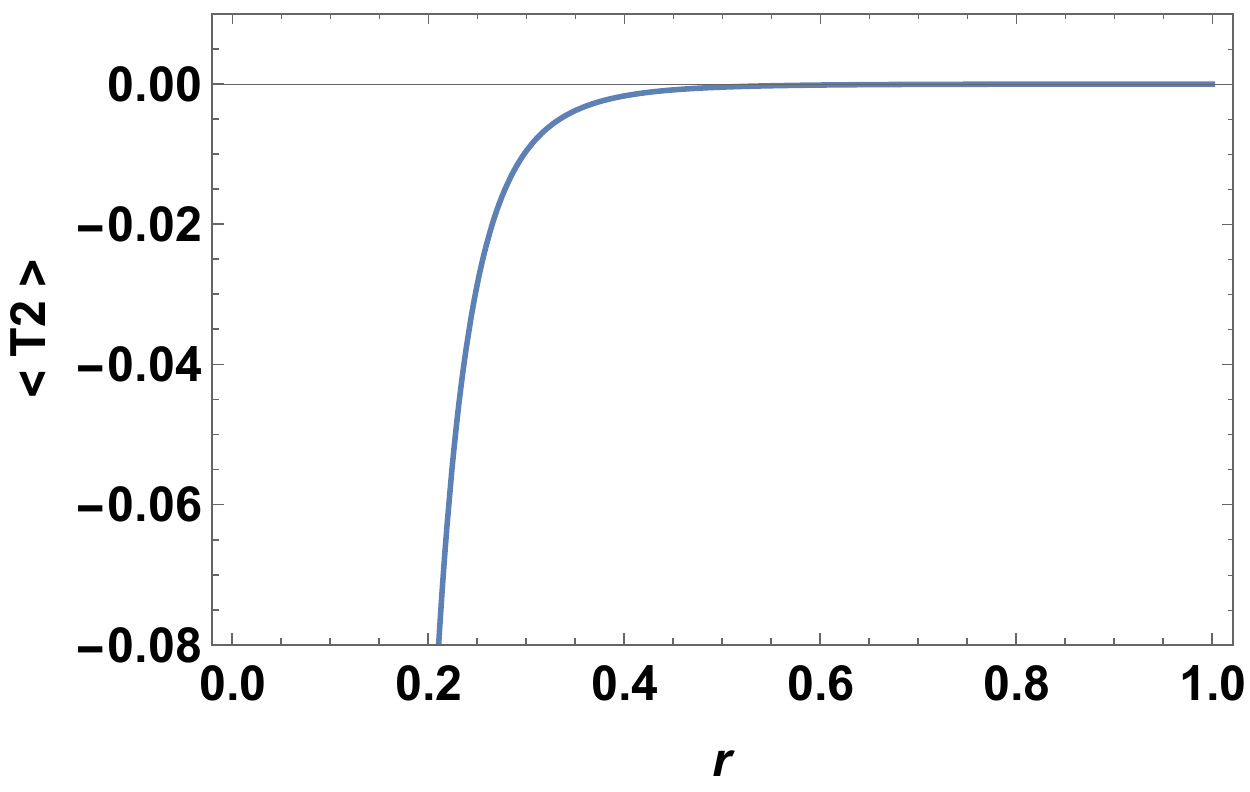}}
\caption{\textit{Dependance on the distance from monopole's core $r$ of the re-scaled angular component of the renormalized stress energy tensor  $\left<T_{2}\right>=96\pi^{2}\mu^{2}\left <T_{\theta}^{\theta}\right >_{ren}=96\pi^{2}\mu^{2}\left <T_{\phi}^{\phi}\right >_{ren}$ for massive scalar field in the pointlike global monopole spacetime. The values of the parameters used in the calculations are $1 - \alpha^2 =10^{-5}$ and $\xi=0 (top \ plot)$ and $\xi=\frac{1}{6} (bottom \ plot)$}.} \label{f9}
\end{figure}

Now substituting (\ref{metric}) in (\ref{SET}) we can obtain the radial and angular components of the renormalized stress energy tensor for the massive scalar field with arbitrary coupling parameter $\xi$. For the radial component we obtain the very simple result
\begin{equation}
\left<T_{r}^{r}\right>_{ren}=\frac{\left(1-{\alpha}^{2}\right)}{10080 \, \pi^{2}\mu^{2}r^{6}}\sum_{k=0}^{3}Q_{k}(\alpha)\xi^{k}
\label{Tradial}
\end{equation}
where
\begin{equation}
Q_{0}(\alpha)=4+67 \alpha^{2}\left(1+{\alpha}^{2}\right)
\end{equation}
\begin{equation}
Q_{1}(\alpha)=-\left[42+\alpha^{2}\left(336\alpha^{2}-966\right)\right]
\end{equation}
\begin{equation}
Q_{2}(\alpha)=210+\alpha^{2}\left(5040-1890\alpha^{2}\right)
\end{equation}
and
\begin{equation}
Q_{3}(\alpha)=-\left[420+\alpha^{2}\left(9660\alpha^{2}-9240\right)\right]
\end{equation}

For the angular components of the renormalized stress energy tensor we obtain
\begin{equation}
\left<T_{\theta}^{\theta}\right>_{ren}=\left<T_{\varphi}^{\varphi}\right>_{ren}=-\frac{1}{2}\left<T_{r}^{r}\right>_{ren}
\label{Tangular}
\end{equation}

In Figure \ref{f4} we show the dependance on the coupling constant $\xi$ of the re-scaled radial component of the renormalized stress energy tensor $\left<T_{1}\right>=96\pi^{2}\mu^{2}\left <T_{r}^{r}\right >_{ren}$ for massive scalar field in the pointlike global monopole spacetime with parameter $1 - \alpha^2 =10^{-5}$ at a fixed distance from monopole's core.

We can observe that $\left<T_{1}\right>$ decreases with the increase of the coupling constant until it reach its minimum value at $\xi=0.2$, then increasing its value for $\xi\geq 0.2$. For all values of the coupling constant this magnitude remains positive, and this behaviour is independent of the value of the distance $r$ from monopole's core, as we can see in Figure \ref{f5}, where we show the dependance of $\left<T_{1}\right>$ as a function of $\xi$ and $r$. Also we observe that $\left<T_{1}\right>$ tends to zero at large distances from  monopole's core. For minimal and conformal coupling Figure \ref{f6} shows this general behaviour.

As (\ref{Tangular}) shows, the angular components of the renormalized stress tensor for the massive scalar field with arbitrary coupling to gravity on global monopole spacetime is proportional, with opposite sign, to the radial component. For this reason it is expected that  $\left<T_{2}\right>$ increases with the increase of the coupling constant until it reach its maximum value at $\xi=0.2$, then decreasing its value for $\xi\geq 0.2$, and remaining negative for all values of the coupling constant, a behaviour independent of the distance $r$ from monopole's core.

This facts can be observed Figures \ref{f7} to \ref{f9}, where we show the dependance of the re-scaled angular component $\left<T_{2}\right>=96\pi^{2}\mu^{2}\left <T_{\theta}^{\theta}\right >_{ren}$ as a function of $\xi$, as function of $\xi$ and $r$, and finally as a function of $r$ for minimal and conformal coupling, respectively. Also we observe that $\left<T_{2}\right>\rightarrow 0$ as $r\rightarrow \infty$, as we expected.

If we defined as usual the energy density of the quantum field as
\begin{equation}
\rho=-\left<T_{t}^{t}\right>_{ren}
\label{eden}
\end{equation}
the above results indicate that, for the massive scalar field in the spacetime of a pointlike global monopole, this magnitude is negative for all values of $r$ if the coupling constant is on the interval $\xi\in\left[0.17, \ 0.23\right]$, and positive for all values of $r$ outside this interval, which includes the minimal and conformally coupled case.

The principal pressures related with the diagonal components of the renormalized stress energy tensor can be defined, in the usual way, by
\begin{equation}
p_{r}=-\left<T_{r}^{r}\right>_{ren}
\end{equation}
for the radial pressure and
\begin{equation}
p_{\theta}=p_{\phi}=p=\left<T_{\theta}^{\theta}\right>_{ren}=\left<T_{\varphi}^{\varphi}\right>_{ren}
\end{equation}
for the angular ones.

As we can see from the results discussed above, the radial pressure is negative for all values of the coupling constant, a behaviour independent of the distance from the monopole's core. The angular pressures remains negative for all values of $\xi$ and $r$ too.

The above facts are interesting in relation with possible violations of energy conditions by the quantized massive scalar field in the global monopole background. As all the information of the quantum field in this background is encoded in the components of the stress energy tensor, one can gain better information on the nature of this quantized field analysing the fulfillment or not of the various energy conditions that can be considered in this case.

Energy conditions are restrictions that the components of the stress energy tensor of matter fields do satisfy to be in some sense reasonable types of matter, and are important in the proofs of various theorems, such as those concerned with singularities, topological censorship and positivity of mass.

The pointwise energy conditions, in the case of a spherically symmetric spacetime, can be summarized as follows \cite{visser,matyjasekEC}:

\textbf{Null energy condition (NEC)}: A matter field satisfies the pointwise NEC if, for any null vector $k^{\mu}$, we have that $T_{\mu \nu}k^{\mu}k^{\nu}\geq 0$. In terms of the diagonal components of the stress energy tensor, the above condition is equivalent to the restrictions $\rho-p_{r}\geq 0$ and $\rho+p\geq 0$.

\textbf{Weak energy condition (WEC)}: The pointwise WEC is satisfied by a matter field if, for any timelike vector $V^{\mu}$ we have $T_{\mu \nu}V^{\mu}V^{\nu}\geq 0$, which is equivalent to the restrictions $\rho+p_{i}\geq 0$ and $\rho\geq 0$. Then, the WEC is equivalent to the NEC with the constraint $\rho\geq 0$ added.

\textbf{Strong energy condition (SEC)}: A matter field satisfies the SEC if $\rho+p_{i}\geq 0$ and $\rho+\sum_{i}p_{i}\geq 0$, which is equivalent to NEC with the constraint $\rho-p_{r}+2p\geq 0$ added.

\textbf{Dominant energy condition (DEC)}: It is satisfied by a matter field whose locally measured energy density is positive and the energy flux is timelike or null. This is equivalent to the restriction $\rho\geq 0$ and $-\rho\leq p_{i}\leq \rho$.

As it is difficult to analyze the fulfillment of the above pointwise energy conditions in the case of a quantum massive scalar field on global monopole spacetime for the case of arbitrary $\xi$, we will restrict our analysis to the physical cases of minimal and conformal couplings.

Re-scaling the energy density and pressures of the quantum massive scalar field as $\varrho=96\pi^{2}\mu^{2} \rho$ and $\textrm{p}_{i}=96\pi^{2}\mu^{2} p_{i}$, the results obtained for the renormalized stress energy tensor indicates that, in the minimally coupled case, we have the relations $\varrho>0$ and
\begin{equation}
\varrho-\textrm{p}_{r}=\frac{1.59 \cdot10^{-4}}{r^{6}}
\end{equation}
\begin{equation}
\varrho+\textrm{p}=-\frac{3.71 \cdot10^{-5}}{r^{6}}
\end{equation}
\begin{equation}
\varrho-\textrm{p}_{r}+2\textrm{p}=-\frac{1.03 \cdot10^{-4}}{r^{6}}
\end{equation}

Also, in the conformally coupled case we have  $\varrho>0$ and
\begin{equation}
\varrho-\textrm{p}_{r}=\frac{4.44 \cdot10^{-6}}{r^{6}}
\end{equation}
\begin{equation}
\varrho+\textrm{p}=-\frac{6.03 \cdot10^{-6}}{r^{6}}
\end{equation}
\begin{equation}
\varrho-\textrm{p}_{r}+2\textrm{p}=-\frac{9.52 \cdot10^{-6}}{r^{6}}
\end{equation}

The above results show that, for both minimal and conformally coupled cases, we have $\varrho > 0$, $\varrho-\textrm{p}_{r} > 0$, $\varrho+\textrm{p} < 0$ and $\varrho-\textrm{p}_{r}+2\textrm{p} < 0$ for all values of $r$. Then, outside the pointlike global monopole's core, all the pointwise energy conditions are violated by the quantum massive scalar field.
\section{Conclussions}

In this paper we construct the approximate renormalized stress energy tensor $\left<T_{\mu \nu}\right>_{ren}$, for a quantum massive scalar field with arbitrary coupling to gravity in the space-time of a pointlike global monopole. Using the leading term in the Schwinger-DeWitt expansion for the Green's function associated with the Klein-Gordon dynamical operator, we find analytical expressions for the one-loop effective action, as an expansion in powers of inverse squared mass of the field. Then, by functional differentiation of the effective action with respect to the metric tensor, we find an analytic expression for $\left<T_{\mu \nu}\right>_{ren}$, valid for a generic spacetime background.

The results obtained for the renormalized stress energy tensor of the quantized massive scalar field in the global monopole background shows that, for all values of the distance $r$ from monopole's core, the quantum massive scalar field violates all the pointwise energy conditions for the minimal and conformally coupled cases.

Our calculations is a sequel of previous work in which, using the Schwinger-DeWitt proper time formalism, we study of vacuum polarization of massive fields in pointlike global monopole's background constructing the analytic formula for the renormalized vacuum expectation value of the square of the field amplitude $\left<\phi^{2}\right>$ \cite{owenmonopole1}.

As a check of the result presented in this paper, we can use the known fact that, for conformally coupled scalar field, there exist a relation between the field fluctuation of the scalar field and the trace anomaly of the stress energy tensor. As showed by Anderson in reference \cite{anderson}, we have for the trace $\left<T_{\nu}^{\nu}\right>$ the expression
\begin{equation}
\left<T_{\nu}^{\nu}\right>=\frac{[a_{2}]}{16 \pi^{2}}-\mu^{2}\left<\phi^{2}\right>
\label{tanomaly}
\end{equation}

In \cite{owenmonopole1} we obtain for the trace of the renormalized stress tensor of the massive scalar field in the pointlike global monopole spacetime the result
\begin{equation}
\left<T_{\nu}^{\nu}\right>=\frac{25 \alpha ^6-21 \alpha ^2-4}{22680 \pi ^2 \mu ^2 r^6}
\label{tanomalymonopole}
\end{equation}

We can easily show that the above result coincides exactly with the trace of $\left<T_{\mu}^{\nu}\right>_{ren}$ obtained using the general results presented in this paper, which is an indication of the validity of the results reported here.

An interesting problem to be addressed is the study of the backreaction effects of the quantized massive fields upon the spacetime geometry around the monopole. Using the analytic expressions for the quantum stress tensor reported here, we can solve perturbatively the semiclassical Einstein's field equations. To this specific problem we will dedicate a future report.

\section*{Acknowledgments}
This work has been supported by TWAS-CONACYT $2017$ fellowship, that allow to the author to do a sabatical leave at Departamento de F\'isica Te\'orica, Divisi\'on de Ciencias e Ingenier\'ias, Universidad de Guanajuato, Campus Le\'on. The author also express his gratitude to Professor Oscar Loaiza Brito, for the support during the research stay at his group, where this work was completed.


\end{document}